\documentclass{elsart}
\usepackage[dvips]{graphicx}
\journal{Physics Letters A}
\begin{document}
\begin{frontmatter}

\title{Numerical Study of Phase Transition in an Exclusion Model with Parallel Dynamics}
\author{Farhad H Jafarpour}
\address{Physics Department, Bu-Ali Sina University, Hamadan,Iran\\
Institute for Studies in Theoretical Physics and Mathematics
(IPM), P.O. Box 19395-5531, Tehran, Iran}
\ead{farhad@sadaf.basu.ac.ir}
\begin{abstract}
A numerical method based on Matrix Product Formalism is proposed
to study the phase transitions and shock formation in the
Asymmetric Simple Exclusion Process with open boundaries and
parallel dynamics. By working in a canonical ensemble, where the
total number of the particles is being fixed, we find that the
model has a rather non-trivial phase diagram consisting of three
different phases which are separated by second-order phase
transition. Shocks may evolve in the system for special values of
the reaction parameters.
\end{abstract}
\end{frontmatter}
\begin{figure}[htbp]
\includegraphics[height=6cm] {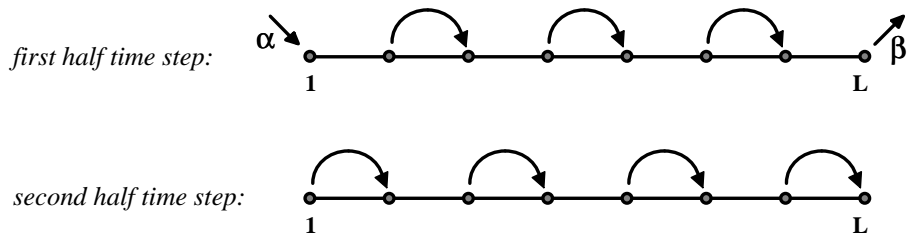}
\caption{\label{fig1} Parallel updating scheme for the ASEP with
open boundary.}
\end{figure}
\begin{figure}[htbp]
\begin{center}
\includegraphics[height=5cm] {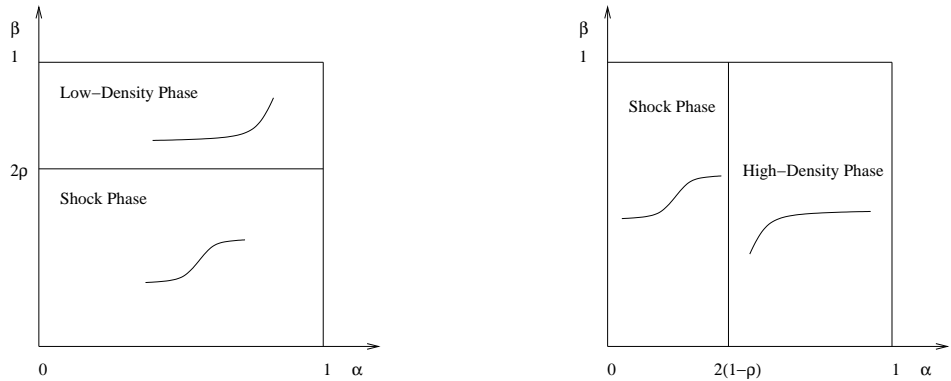}
\caption{\label{fig2} The phase diagrams of the ASEP in canonical
ensemble with parallel dynamics and open boundaries for $\rho <
\frac{1}{2}$ (left) and $\rho > \frac{1}{2}$ (right). The small
curves show the behaviors of the density profile of particles in
each phase.}
\end{center}
\end{figure}
\begin{figure}[htbp]
\begin{center}
\setlength{\unitlength}{1mm}
\begin{picture}(0,0)
\put(-6,47){\makebox{$\scriptstyle \rho(x)$}}
\put(-6,41){\makebox{$\scriptstyle \rho_2 $}}
\put(-6,17){\makebox{$\scriptstyle \rho_1  $}}
\put(75,-2){\makebox{$\scriptstyle x      $}}
\put(32,-2){\makebox{$\scriptstyle x_0    $}}
\end{picture}
\includegraphics[height=5cm] {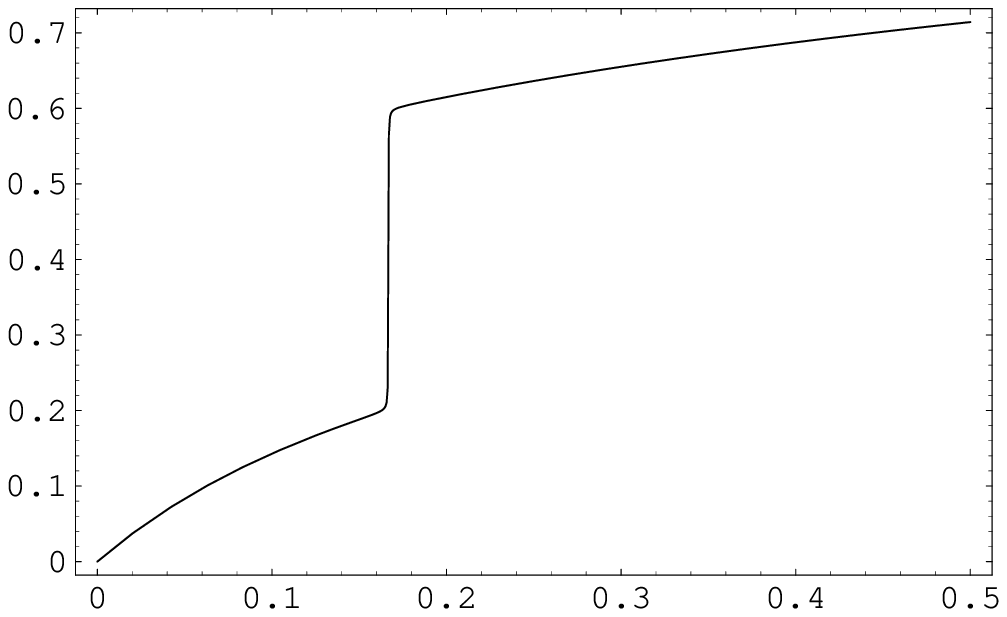}
\caption{\label{fig3} The density of particles $\rho(x)$ as a
function of their fugacity $x$ for $L=15000$, $\alpha=0.8$ and
$\beta=0.4$. The critical densities are
$\rho_1=\frac{\beta}{2}=0.2$ and $\rho_2=1-\frac{\alpha}{2}=0.6$.}
\end{center}
\end{figure}

During the last decade the study of one-dimensional driven lattice
gases has been of increasing interest because besides their
application in different areas of non-equilibrium physics they
show a variety of fascinating properties such as non-equilibrium
phase transitions and spontaneous symmetry breaking \cite{sch1}.
They have also let us study the one-dimensional shocks i.e.
discontinuities in the density of particles on the lattice over a
microscopic region. Depending on the dynamics these models can be
divided into two different classes: models with sequential
(continuous time evolution) and parallel (discrete time evolution)
dynamics. These models can also have open boundaries (where the
particles can enter or leave the lattice) or periodic boundary
conditions (with conservation of the number of particle). \\
Different approaches have been used to study the one-dimensional
out-of-equilibrium models. Among these approaches the Matrix
Product Formalism (MPF) is the one which allows us to study these
models under sequential or parallel dynamics with different
boundary conditions \cite{dehp}. According to the MPF the
stationary probability distribution function of the system can be
written in terms of the matrix element (for the open boundary
case) or trace (for the periodic boundary case) of products of
non-commuting operators. Several models have been studied using
the MPF. In some cases they can be solved exactly; however, in
most of the cases exact solutions can only be found in small
regions of the reaction parameters space. In a recent work
\cite{farhad1} it has been shown that the MPF can be used in order
to numerical study of one-dimensional reaction-diffusion models
with sequential dynamics and conservation of the number of
particles. In the present letter we aim to show that this approach
can also be applied to the numerical study of one-dimensional
reaction-diffusion models with parallel dynamics and conservation
of particles. The Asymmetric Simple Exclusion Process (ASEP) with
parallel dynamics will be considered as an example.\\
In the ASEP with parallel dynamics \footnote{This is sometimes
called sublattice-parallel updating scheme in related
literatures.} classical particles move on a one-dimensional
lattice of length $L$ (which is assumed to be an even number) with
open boundaries. The bulk dynamic is deterministic and consists of
two half time steps. In the first half time step particles at even
sites move one step to the right provided that their rightmost
sites are empty. In this step both the first and the last sites
are also updated. From the first (last) site the particles are
injected (extracted) with the probability $\alpha$ ($\beta$)
provided that the target site is empty (occupied). In the second
half time step only the odd sites are updated and the particles at
these sites move to the right in the same way. The parallel
updating scheme is shown in Fig.~\ref{fig1}.
The ASEP with open boundaries and parallel dynamics has been
originally proposed in \cite{sch2}. Later it was studied using the
MPF in \cite{hin}. They have both worked in the grand canonical
ensemble where the total number of particles in the system is not
conserved. The result is that in the large-$L$ limit the model has
two different phases: a high-density phase for $\alpha > \beta$
and a low-density phase for $\alpha < \beta$. A first-order phase
transition also takes place at the transition point $\alpha=\beta$
where the density profile of particles changes linearly along the
lattice. The linear density profile of particles is related to the
superposition of shocks in the system. In order to study the
shocks one may consider the model on a ring (a lattice with
periodic boundary conditions) either in the presence of a slow
particle \cite{mall,lpk,farhad2}, which is sometimes called an
impurity, or a slow link \cite{hinsan}. Equivalently one can leave
the boundaries open and restrict the total number of particles by
working in a canonical ensemble. In the present letter we adopt
the latter scenario i.e. we leave the boundaries open; however,
the total number of particles on the lattice $M$ (and therefore
their density $\rho=\frac{M}{L}$) is restricted to being a constant. \\
It is shown that the stationary probability distribution function
of the ASEP with parallel dynamics can be written as \cite{hin}
\begin{equation}
\label{MatrixAnsatz}
P(\tau_1,\ldots,\tau_{L}) = \frac{1}{Z_L} \langle W \vert
\prod_{i=1}^{\frac{L}{2}} [\hat{\mathcal{O}}_{2i-1}\mathcal{O}_{2i}]
\vert V \rangle
\end{equation}
in which we have defined
\begin{equation}
\label{operators}
\hat{\mathcal{O}}_i:=\tau_{i} \hat{D} + (1-\tau_{i})\hat{E} \; , \;
\mathcal{O}_i:=\tau_{i} D + (1-\tau_{i})E
\end{equation}
and $\tau_i \in \{0,1\}$ is the occupation number at site $i$. The
denominator in (\ref{MatrixAnsatz}) is a normalization factor. The
matrices ($\hat{D},\hat{E}$) and ($D,E$) are square matrices and
besides the vectors $\vert V \rangle$ and $\langle W \vert$ are
acting in an auxiliary space which might have either a finite or
an infinite dimension. They also stand for the presence of
particles and holes at odd and even sites respectively and should
satisfy the following quadratic algebra
\begin{equation}
\begin{array}{l}
\label{BulkAlgebra}
[E,\hat{E}] = [D,\hat{D}] = 0 \; ,
\; E \hat{D} = [\hat{E},D] \; , \; \hat{D} E = 0 \\
\langle W \vert \hat{E} (1-\alpha) = \langle W \vert E \; , \;
\langle W \vert (\alpha \hat{E} + \hat{D}) = \langle W \vert D \\
(1-\beta) D \vert V \rangle = \hat{D} \vert V \rangle \; , \;
(E+\beta D) \vert V \rangle = \hat{E}\vert V \rangle.
\end{array}
\end{equation}
It has also been shown that (\ref{BulkAlgebra}) has a
two-dimensional representation for $\alpha \neq \beta$ \cite{hin}
\begin{equation}
\begin{array}{l}
\label{BulkRep} \hat{D} = \left( \begin{array}{cccc}
\alpha (1-\beta) & & & 0 \\
-\alpha \beta & & & 0
\end{array} \right) \; , \;
\hat{E} = \left( \begin{array}{cccc}
\alpha \beta & & & 0  \\
\alpha\beta & & & \beta
\end{array} \right) \; ,\;
\vert V \rangle = \left( \begin{array}{c}
1-\beta \\ -\beta \end{array} \right) \\ \\
D = \left( \begin{array}{cccc}
\alpha  & & & 0 \\
-\alpha \beta & & & \alpha \beta
\end{array} \right) \; , \;
E = \left( \begin{array}{cccc}
0 & & & 0  \\
\alpha\beta & & & \beta (1-\alpha)
\end{array} \right) \; , \;
\langle W \vert = (\alpha, 1-\alpha).
\end{array}
\end{equation}
The normalization factor $Z_L$ in (\ref{MatrixAnsatz}), which will
be called the grand canonical partition function for $\alpha
> \beta$ hereafter, can easily be calculated using the fact that
$\sum_{\{ \mathcal C\}}P(\tau_1,\cdots,\tau_L)=1$. It is found
\begin{equation}
\label{GCPF} Z_L=\langle W \vert
[(\hat{E}+\hat{D})(E+D)]^{\frac{L}{2}} \vert V
\rangle=(1-\beta)\alpha^{L+1}-(1-\alpha)\beta^{L+1}.
\end{equation}
For $\alpha < \beta$ the grand canonical partition function should
be defined as $-Z_L$. Let us investigate the phase transitions of
the model in this case using the classical Yang-Lee theory
\cite{yanglee,gross}. One can apply the Yang-Lee theory by finding
the roots of the partition function of the system as a function of
one of its intensive variables. It can readily be seen that in the
thermodynamic limit $L \rightarrow \infty$ the zeros of
(\ref{GCPF}) as a function of $\alpha$ ($\beta$) lie on a circle
of radius $\beta$ ($\alpha$). This predicts a first-order phase
transition at $\alpha=\beta$. Since the representation of the
algebra (\ref{BulkAlgebra}) has a finite dimensional
representation, we expect that the density profile of the
particles has an exponential behavior in both phases $\alpha <
\beta$ and $\alpha > \beta$. This is in agreement with the results
obtained in \cite{sch2,hin} for the phase diagram of the model. As
we mentioned above at the phase transition point one finds a
linear profile for the density of particles on the lattice. This
is related to the superposition of shocks which can be anywhere on
the lattice. In what follows we will investigate the phase diagram
of the model with fixed number of particles by introducing a
canonical partition function as
\begin{equation}
\label{CPF}
Z_{L,M}=\sum_{\{\tau_i=0,1\}}\delta(M-\sum_{i=1}^{L}\tau_i)\langle W \vert
\prod_{i=1}^{\frac{L}{2}} [\hat{\mathcal{O}}_{2i-1}\mathcal{O}_{2i}]
\vert V \rangle
\end{equation}
in which $M$ and $L$ are the number of particles and the length of
the system respectively and $\delta(x)$ is the ordinary Kronecker
delta function $\delta_{x,0}$. The operators $\hat{\mathcal{O}}_i$
and $\mathcal{O}_i$ are also defined in (\ref{operators}). It can
easily be verified that (\ref{CPF}) can be written as
\begin{equation}
\label{GeneralizedCPF}
Z_{L,M}=\langle W \vert Coefficient[(\hat{C}C)^{\frac{L}{2}},M]
\vert V \rangle
\end{equation}
in which we have defined
\begin{equation}
\label{CC}
\begin{array}{c}
\hat{C}:= \hat{E}+x\hat{D} \\ C:=E+xD
\end{array}
\end{equation}
where $x$ is a free parameter and $Coefficient[Expr,n]$ gives the
coefficient of $x^n$ in the expression $Expr$. Note that since
$Expr$ is a matrix then $Coefficient[Expr,n]$ gives the
coefficient $x^n$ of each of its entries. A similar formula has
already been introduced in \cite{farhad1} which can be used for
the numerical calculation of the canonical partition function of
one-dimensional reaction-diffusion models with sequential
dynamics. The formula (\ref{GeneralizedCPF}) can be useful for
numerical studying of phase transitions in out-of-equilibrium
systems with parallel dynamics. For the ASEP with parallel
dynamics we will calculate (\ref{GeneralizedCPF}) as a function of
$\alpha$ and $\beta$ for finite values of $L$ and $M$ by using the
representation of its quadratic algebra (\ref{BulkRep}).\\
Now we investigate the density profile of the particles on the
lattice which is defined as
\begin{equation}
\label{DP}
\rho(i)=\frac{\sum_{\{\mathcal C \}}\tau_i
P(\tau_i,\cdots,\tau_L)}{\sum_{\{\mathcal C
\}}P(\tau_i,\cdots,\tau_L)}
\end{equation}
where $\mathcal{C}$ is any configuration with a fixed number of
particles and $P(\tau_i,\cdots,\tau_L)$ is given by
(\ref{MatrixAnsatz}). The formula (\ref{DP}) can be written in
terms of the operators $\hat{D},\hat{E},D,E$ and the vectors
$\vert V \rangle$ and $\langle W \vert$; however, since the
operators for the existence of particles at even and odd sites are
not the same we find different expressions for the density of
particles at odd and even sites. It can easily be seen that the
density of particles at odd and even sites can be written as
\begin{equation}
\begin{array}{c}
\label{GDPodd}
\rho(2i-1)= \frac{1}{Z_{L,M}}\sum_{k=0}^{M-1} \\
\langle W \vert Coefficient[(\hat{C}C)^{i-1},k] \hat{D}
\;Coefficient[C(\hat{C}C)^{\frac{L}{2}-i},M-k-1] \vert V
\rangle\end{array}\end{equation}and \begin{equation}
\begin{array}{c}
\label{GDPeven}
\rho(2i)=\frac{1}{Z_{L,M}}\sum_{k=0}^{M-1} \\
\langle W \vert Coefficient[(\hat{C}C)^{i-1}\hat{C},k] D
Coefficient[(\hat{C}C)^{\frac{L}{2}-i},M-k-1] \vert V \rangle
\end{array}
\end{equation}
respectively where $i=1,\cdots,\frac{L}{2}$. The formulas
(\ref{GeneralizedCPF}), (\ref{GDPodd}) and (\ref{GDPeven}) are
quite general and can be used for any one-dimensional
reaction-diffusion model with open boundaries, parallel
dynamics and only one class of particles. \\
In order to study the phase transitions of the ASEP in canonical
ensemble with open boundaries and parallel dynamics, we use the
classical Yang-Lee theory by investigating the zeros of
(\ref{GeneralizedCPF}) in the complex plane of both $\alpha$ and
$\beta$ for different values of $L$ and $M$. The particle
concentration at odd and even sites will also be calculated from
(\ref{GDPodd}) and (\ref{GDPeven}) in each phase. Our numerical
calculations show that in the thermodynamic limit ($L\rightarrow
\infty,M\rightarrow \infty, \rho=\frac{M}{L}$) the model has two
different phase diagrams depending on the density of particles on
the lattice $\rho$: For $\rho < \frac{1}{2}$ the phase diagram of
the model consists of a low-density and a shock phase which are
separated by a second-order phase transition at $\beta_c=2\rho$.
The reason that the phase transition is of a second-order is that
the zeros of (\ref{GeneralizedCPF}) as a function of $\beta$,
approach the positive real-$\beta$ axis at an angle
$\frac{\pi}{4}$ \cite{be}. The canonical partition function
(\ref{GeneralizedCPF}) as a function of $\alpha$, does not have
any real and positive root smaller than one in this case. In the
low-density phase $\beta > \beta_c$ the density of particles at
both even and odd sites changes exponentially along the lattice;
however, in the shock phase $\beta < \beta_c$ as we move from the
first to the last site of the lattice the particle concentration
changes abruptly from $\beta$ to $1$ for even sites and from $0$
to $1-\alpha$ for odd sites. For $\rho > \frac{1}{2}$ the phase
diagram of the model consists of a high-density and a shock phase
which are separated by a second-order phase transition at
$\alpha_c=2(1-\rho)$. As for the case $\rho < \frac{1}{2}$ the
zeros of (\ref{GeneralizedCPF}) as a function of $\alpha$,
approach the positive real-$\alpha$ axis at an angle
$\frac{\pi}{4}$ and therefore the phase transition is of
second-order \cite{be}. The canonical partition function
(\ref{GeneralizedCPF}) as a function of $\beta$ does not have any
real and positive root smaller than one in this case. In the
high-density phase $\alpha > \alpha_c$ the density of particles at
both even and odd sites changes exponentially; however, the shock
phase in this case takes place at $\alpha < \alpha_c$ and the
structure of the particles concentrations in quite similar to the
shock phase for $\rho < \frac{1}{2}$. The phase diagrams for both
$\rho < \frac{1}{2}$ and $\rho > \frac{1}{2}$ cases are shown in
Fig.~\ref{fig2}.
At $\rho=\frac{1}{2}$ the phase diagram of the model consists of
only a shock phase independent of the values of $\alpha$ and
$\beta$. For fixed values of the injection and extraction
probabilities ($\alpha_0,\beta_0$) the phase diagram of the model
consists of three phases which are determined by the density of
particles $\rho$. For $0 < \rho < \frac{\beta_0}{2}$ we are in the
low-density phase in which the density of particles changes
exponentially along the chain. For $\frac{\beta_0}{2} < \rho <
1-\frac{\alpha_0}{2}$ we are in the shock phase. Finally, for
$1-\frac{\alpha_0}{2} < \rho < 1$ we are in the high-density phase
in which the density profile of particles is an exponential function.\\
The phase diagram of the ASEP in canonical ensemble with parallel
dynamics and open boundaries can also be obtained by studying the
grand canonical partition function of this model defined as
\begin{equation}
Z_L(x)=\langle W \vert (\hat{C}C)^{\frac{L}{2}} \vert V \rangle
\end{equation}
in which $\hat{C}$ and $C$ are given by (\ref{CC}) and $x$ plays
the role of the fugacity of particles. This can easily be
calculated and we find
\begin{equation}
\begin{array}{c}
Z_L(x)=(1-\beta)\alpha^{L+1}x^{\frac{L}{2}}(\beta+x(1-\beta))^{\frac{L}{2}}
-(1-\alpha)\beta^{L+1}((1-\alpha)+x\alpha)^{\frac{L}{2}}\\
+\alpha(1-\alpha)\beta(1-\beta)(1-x)(x\alpha+\beta)\\ \sum_{i=0}^{\frac{L}{2}-1}
x^{\frac{L}{2}-i-1}\alpha^{L-2i-2}\beta^{2i}(\beta+x(1-\beta))^{\frac{L}{2}-i-1}
((1-\alpha)+x\alpha)^{i}.
\end{array}
\end{equation}
Now let us study the total density of particles as a function of
the fugacity $x$
\begin{equation}
\rho(x)=\frac{x}{L}\frac{\partial}{\partial x} \log Z_L(x).
\end{equation}
In Fig.~\ref{fig3} we have plotted $\rho(x)$ as a function of $x$
for $L=15000$, $\alpha=0.8$ and $\beta=0.4$.  As can be seen, the
density of particles is an increasing function of $x$ up to a
critical point $x_0$ where a finite discontinuity takes place.
Above the critical point $x_0$ the density increases very slowly
and remains smaller than unity unless $\alpha \rightarrow 0$. Our
numerical calculations show that the discontinuity starts from
$\rho_1=\frac{\beta}{2}$ and ends at $\rho_2=1-\frac{\alpha}{2}$
(see Fig.~\ref{fig3}). This is quite in agreement with the picture
that we got for the phase diagram of this model.
For $x < x_0$ and $x > x_0$ the density of particles is determined
by their fugacity; however, at $x_0$ the fugacity does not fix the
density. Here is where we have shocks in the system. This
phenomenon has already been observed in other models where the
Bose-Einstein condensation takes place and the conservation of the
number of particles is broken \cite{farhad3}.\\
In this letter we have investigated the phase transitions and
shock formation in the ASEP with open boundaries and parallel
dynamics in a canonical ensemble where the total number of
particles is kept fixed. We have found that the phase diagram of
the model depends of the value of the density of particles on the
system. The system can be in any of its three accessible phases:
the low-density phase, the high-density phase and the shock-phase.
In the shock-phase, the shock position is fixed and determined by
the number of particles while in the hight-density and the
low-density phases the density profile of the particles at both
even and odd sites have exponential behaviors. Since the
representation of the associated algebra (\ref{BulkAlgebra}) is
finite dimensional, we do not expect the density profile of
particles on the lattice to have algebraic behavior \cite{hkp}.

\end{document}